\documentstyle[aps,epsfig]{revtex}

\renewcommand{\baselinestretch}{1.1}

\newcommand{\bc}{\begin{center}}
\newcommand{\ec}{\end{center}}
\newcommand{\bmi}{\begin{minipage}}
\newcommand{\emi}{\end{minipage}}
\newcommand{\bdi}{\begin{displaymath}}
\newcommand{\edi}{\end{displaymath}}
\newcommand{\bit}{\begin{itemize}}
\newcommand{\eit}{\end{itemize}}

\newcommand{\Ne}{N\'{e}el}

\newcommand{\bi}{\begin{itemize}}
\newcommand{\ei}{\end{itemize}}
\newcommand{\be}{\begin{equation}}
\newcommand{\ee}{\end{equation}}

\newcommand{\bea}{\begin{eqnarray}}
\newcommand{\eea}{\end{eqnarray}}   

\newcommand{\bdm}{\begin{displaymath}}
\newcommand{\edm}{\end{displaymath}}
\newcommand{\beas}{\begin{eqnarray*}} 
\newcommand{\eeas}{\end{eqnarray*}}
\newcommand{\bite}{\begin{itemize}}
\newcommand{\enite}{\end{itemize}}
\newcommand{\ben}{\begin{enumerate}}
\newcommand{\een}{\end{enumerate}}

\newcommand{\kag}{{\it kagom{\'e}{\ }}}
\newcommand{\sez}{\mbox{$s={1\over2}$}}
\newcommand{\sen}{\mbox{$s=1$}}
\newcommand{\sdz}{\mbox{$s={3\over2}$}}

\begin{document}

\author{A.~Voigt}
\address{Center for Simulational Physics,
Department of Physics and Astronomy,
University of Georgia, Athens GA 30605 USA
\footnote{Tel. +1-706-542-3867, Fax +1-706-542-2492,
E-mail: andreas@physast.uga.edu}
}

\author{J.~Richter}

\address{Institut f\"ur Theoretische Physik,
Otto-von-Guericke-Universit\"at Magdeburg, 
PF 4120, 39106 Magdeburg, Germany}

\author{P.~Tomczak}

\address{Uniwersytet im. Adama Mickiewicza, Wydzia{\l} Fizyki,
ul.Umultowska 85, 61-614 Pozna{\'n}, Poland}

\title{The quantum Heisenberg antiferromagnet on the Sierpi{\'n}ski Gasket:
An exact diagonalization study}

\maketitle
\bibliographystyle{prsty}

\bc
(\today)\\
\ec

\begin{abstract} 
We present an exact diagonalization study of the quantum Heisenberg
antiferromagnet on the fractal Sierpi{\'n}ski gasket for spin quantum
numbers \sez, $s=1$ and $s={3\over2}$.  Since the fractal dimension of
the Sierpi{\'n}ski gasket is between one and two we compare the results
with corresponding data of one- and two-dimensional systems. By
analyzing the ground-state energy, the low-lying spectrum, the
spin-spin correlation and the low-temperature thermodynamics we find
arguments, that the Heisenberg antiferromagnet on the Sierpi{\'n}ski
gasket is probably disordered not only in the extreme quantum case
$s={1\over2}$ but also for $s=1$ and $s={3\over2}$. Moreover, in
contrast to the one-dimensional chain we do not find a distinct
behavior between the half-integer and integer-spin Heisenberg models on
the Sierpi{\'n}ski gasket.  We conclude that magnetic disorder may
appear due to the interplay of frustration and strong quantum
fluctuations in this spin system with spatial dimension between one and
two.
\end{abstract}

\bc
PACS numbers: 75.10.Jm, 75.50.Ee, 75.40.Mg
\ec

\section{Introduction}

For one-dimensional (1d) and two-dimensional (2d) quantum Heisenberg
antiferromagnets the question about the existence of magnetic
long-range order (LRO) in the ground state (GS) is of great importance. 
Since only in some special cases analytical solutions exist the
numerical methods like quantum Monte Carlo and exact diagonalization
play an important role for this many-body problem. For the 1d linear
chain with \sez \ the famous Bethe-Ansatz \cite{bethe31,hulthen38}
provides the analytic solution: In the GS the spin-spin correlation
exhibits a power-law decay to zero and consequently there is no
magnetic LRO. For the 1d Heisenberg antiferromagnet (HAF) with higher
spin quantum number $s > {1\over2}$ no analytic solutions are
available, but comprehensive studies have shown that the ground state
exhibits no LRO. A special aspect of the antiferromagnetic chain is the
fundamental difference between half-integer and integer spin HAF as
conjectured by Haldane \cite{haldane83}. While for the half-integer
spin chain (\sez$,{3\over2},\ldots$) the spin-spin correlation in the
GS decays to zero according to a power-law and there is no gap between
the singlet GS and the first triplet excitation (i.e. the GS is
critical), for the integer spin chain (\sen$,2,\ldots$) the spin-spin
correlation in the GS exponentially decays to zero and there is a
finite excitation gap, the so-called spin gap (i.e. the GS is
disordered).  In contrast to the lack of {\Ne} LRO for 1d
antiferromagnets, most of the 2d antiferromagnets show {\Ne} LRO in the
GS.  This holds for the square lattice, the honeycomb lattice as well
as for the frustrated triangular lattice. All these systems show a
\Ne-like magnetic order resembling the classical {\Ne} state but with a
reduced sublattice magnetization due to the quantum fluctuations. The
general features of the magnetic ordering in these 2d antiferromagnets
does not depend on the spin quantum number $s$, i.e. no qualitative
difference between half-integer and integer spin has been observed.

In this paper we investigate the quantum HAF on the Sierpi{\'n}ski
gasket, a fractal self-similar structure with geometrical
inhomogeneities (see, e.g. \cite{gefen,tomczak96co}).  The first
motivation to investigate this problem is due to the reduced fractal
lattice dimension $d=\ln(3)/\ln(2)$ of the Sierpi{\'n}ski gasket.
Consequently, the question arises whether the quantum
fluctuations in a system of dimension between 1 and 2 are 
strong enough to destroy the antiferromagnetic LRO.
For the \sez {\ } HAF this problem has been considered recently
\cite{tomczak96co,tomczak96prb,voigt98jmmm}.  It has been suggested
that the quantum fluctuations indeed destroy the antiferromagnetic LRO
and the GS is disordered.
However, from these considerations for \sez {\ } a second interesting
question arises: What happens if the spin quantum number $s$ is
enlarged? Is there a basic difference between half-integer and integer
spin HAF like in one dimension or is the influence of the spin quantum
number on the magnetic ordering of less importance like in two
dimensions? The investigation of this question is the main issue of the
present paper.

The HAF on the Sierpi{\'n}ski gasket represents a complex quantum many
body problem with a nontranslational invariant arrangement of sites and
with magnetic frustration.  Hence, some of the standard methods like
spin-wave theory (which suffers from the lack of translational
invariance) or quantum Monte-Carlo (which suffers from the minus sign
problem due to frustration) are not applicable. However, the exact
diagonalization of small finite systems seems to be particularly
appropriate, since it allows a direct comparison of the HAF with
different spin quantum numbers.  This method has been applied
successfully to various problems in quantum spin magnetism, e.g. the
frustrated ferrimagnet in 1d \cite{ivrischo98}, the {\kag} lattice in
2d \cite{lecheminant97,sindzingre00,hida00}, or the HAF on the
body-centered cubic lattice in 3d \cite{betts98}. Though due to the
limitations to small size conclusions for the thermodynamic limit have
to be drawn with particular care, even the problem of magnetic LRO in
the 2d HAF has been addressed successfully (see, e.g.
\cite{oitmaa78b,schulz92,deutscher93}).

In what follows  we will discuss the GS energy, the low-energy
spectrum, the spin-spin correlation in the GS as well as the
low-temperature specific heat for $s={1\over2},1,{3\over2}$ for the HAF
on the Sierpi{\'n}ski gasket mainly for the gasket with $N=15$ sites
shown in Fig.\ref{fig1}.  In accordance with earlier studies on other
frustrated spin systems \cite{schulz92,dagotto89,richter93} our
investigations of the \sez {\ } case has illustrated
\cite{tomczak96co,voigt98jmmm} that despite of the smallness of the
system under consideration even the Sierpi{\'n}ski gasket with $N=15$
seems to cover some important features of the HAF on the infinite
gasket namely the fast (presumably exponential) decay of the spin-spin
correlation and the peculiar low-temperature specific heat behavior.

\section{The model}

We consider the usual HAF
\be
\hat H = J \sum_{\langle  i,j \rangle} {\vec s}_i  {\vec s}_j 
\ee
with antiferromagnetic exchange $J=1$ between nearest neighbors on the
Sierpi{\'n}ski gasket. The parameters of the system are the spin
quantum number $({\vec s}_i)^2 = s(s+1)$
($s={1\over2},1,{3\over2}$) and the size of the system $N$.
Since the Heisenberg Hamiltonian commutes with the square of total spin
${\vec S}^2$, ${\vec S}=\sum_i {\vec s}_i$, each eigen state of $\hat H$
belongs to a certain subspace of the Hilbert space with fixed quantum
number $S$ of the total spin (${\vec S}^2=S(S+1)$).  To calculate the
GS and low-lying excitations of $\hat H$ for $N=15$ sites we use the Lanczos
technique.

\begin{figure}[ht]
\centerline{
\epsfig{file=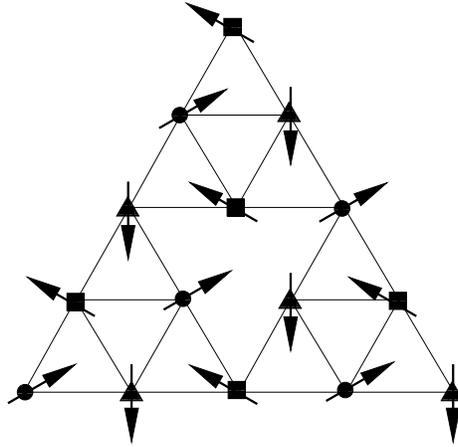,scale=0.45,angle=0}}
\caption{The classical ground state configuration of the HAF on 
the Sierpi{\'n}ski gasket with N=15: 3-sublattices A (circles), B 
(squares) and C (triangles) with an angle of 120$^\circ$ between 
neighboring spins.}
\label{fig1}
\end{figure}

The Hausdorff dimension of this fractal lattice is $d_f={\ln(3) \over
\ln(2)} \approx 1.58$. The number of spins on the gasket is given by
$N={1\over2}(3^n+3)$ with $n=1,2,3,\ldots$.  The ground state of the
classical model (spin quantum number $s=\infty$), where the spins
${\vec s}_i$ effectively can be considered as classical vectors, is the
same as for the HAF on the triangular lattice, namely a {\Ne} state
with 3 sublattices and an angle of $120^\circ$ between neighboring spins. 
This classical ground state is illustrated in Fig.\ref{fig1}.

\section{Ground state energy and low-lying excitations}
\label{ener}

Since the low-lying excitations of a spin system contain signatures of
the magnetic ordering \cite{bernu92,bernu94,lhuillier00} we discuss in
this section the low-energy spectrum of finite systems. First we
compare in Table \ref{eb_vgl} the GS energy per bond $e_b=E/N_b$ (i.e.
the averaged nearest-neighbor spin-spin correlation) of the linear chain 
and of some 2d lattices with the GS energy of the Sierpi{\'n}ski gasket.

\bc
\begin{table}
\renewcommand{\baselinestretch}{0.8}
\huge \normalsize \small
\begin{minipage}[t]{13cm}
\begin{tabular}[t]{cllccccc}
 \\
 d & Lattice & N & z & Frustration & $s={1\over2}$ &  $s=1$ &  $s={3\over2}$\\
 \\
\hline\\ 
 1                 & linear chain        & 12 & 2 & no  & -0.448949  & -1.417120  & -2.844275  \\
 & \dotfill \\
 $ln(3)\over ln(2)$ & Sierpinski gasket  & 15 & 4 & yes & -0.231181  & -0.733279  & -1.466517  \\ 
 & \dotfill \\
 2                 & honeycomb lattice   & 12 & 3 & no  & -0.385048  & -1.264919  & -2.643583  \\
                   & square lattice      & 10 & 4 & no  & -0.365003  & -1.227931  & -2.590695  \\
                   & \kag lattice        & 12 & 4 & yes & -0.226869  & -0.734203  & -1.462779  \\ 
                   & triangular lattice  & 12 & 6 & yes & -0.203443  & -0.653036  & -1.354524  \\ 
\end{tabular}
\end{minipage}
\vspace*{0.3cm}
\caption{Ground state energies per bond $e_b$ of different small lattices 
vs. spin quantum number (d - dimension, N - number of spins, z - number of 
nearest neighbors).
\label{eb_vgl}
}
\end{table}
\ec

We see a clear distinction between frustrated and non-frustrated
systems. For all spin quantum numbers $s$ considered the absolute value
of $e_b$ for non-frustrated systems is about twice as large as for
frustrated ones.  We notice, that for $s \to \infty$ we have exactly a
factor of two between the GS energy per bond of frustrated and
non-frustrated systems, which corresponds to the 120$^\circ$ angle between
neighboring spins. We conclude that the averaged 
nearest-neighbor spin-spin
correlation of the HAF on the Sierpi{\'n}ski gasket is rather close to
that of the triangular and {\kag} lattice.

Next we consider the low-lying excitations of the HAF on the
Sierpi{\'n}ski gasket ($N=15$) classified by the total spin $S$.

\begin{table}
\renewcommand{\baselinestretch}{0.8}
\huge \normalsize \small
\bc
\begin{minipage}[t]{14cm}
\begin{minipage}[t]{4.5cm}

\center{{\large \bf s=$1\over2$}}
\vspace*{1em}

\begin{tabular}[t]{rcc}
\\
     $S$      &    $e_b$   & D \\
\\
\hline 
\\
  $1\over2$ &   -0.231181  & 2 \\ 
            &   -0.229461  & 1 \\ 
            &   -0.229424  & 2 \\ 
            &   -0.222356  & 2 \\ 
            &   -0.221141  & 1 \\ 
            &   -0.220616  & 2 \\ 
            &   -0.219649  & 1 \\ 
            &   -0.219604  & 1 \\ 
  $\dots$   &              &   \\ 
  $3\over2$ &   -0.219233  & 1 \\ 
            &   -0.216872  & 1 \\ 
            &   -0.214222  & 2 \\ 
            &   -0.212654  & 1 \\ 
            &   -0.212386  & 2 \\ 
            &   -0.209587  & 1 \\ 
  $\dots$   &              &   \\ 
  $5\over2$ &   -0.186349  & 1 \\ 
            &   -0.185978  & 2 \\ 
            &   -0.183977  & 1 \\ 
            &   -0.183410  & 2 \\ 
            &   -0.179908  & 2 \\ 
            &   -0.179669  & 1 \\ 
  $\dots$   &              &   \\ 
  $7\over2$ &   -0.136306  & 2 \\ 
            &   -0.132661  & 1 \\ 
            &   -0.127097  & 2 \\ 
            &   -0.126221  & 1 \\ 
            &   -0.118484  & 1 \\ 
            &   -0.115478  & 2 \\ 
\end{tabular} 
\end{minipage}
\hfill
\begin{minipage}[t]{4.5cm}

\center{{\large \bf s=$1$}}
\vspace*{1.3em}

\begin{tabular}[t]{rcc}
\\
  $S$      &    $e_b$   & D \\
\\
\hline 
\\
  0         &   -0.733279    & 1 \\ 
            &   -0.724241    & 1 \\ 
            &   -0.723842    & 2 \\ 
            &   -0.721361    & 1 \\ 
            &   -0.713751    & 1 \\ 
            &   -0.710664    & 2 \\ 
%            &  -0.706843  9  & 2 \\ 
%            &  -0.705263  7  & 2 \\ 
%            &  -0.699512  1  & 2 \\ 
%            &  -0.696080  2  & 2 \\ 
%            &  -0.692286  7  & 2 \\ 
%            &  -0.688642  8  & 2 \\ 
%            &  -0.682392  3  & 2 \\ 
%            &  -0.677504  4  & 2 \\ 
%            &  -0.671494  5  & 2 \\ 
  $\dots$   &                &   \\ 
  1         &   -0.724201    & 2 \\ 
            &   -0.718846    & 1 \\ 
            &   -0.718752    & 2 \\ 
            &   -0.715339    & 1 \\ 
            &   -0.714232    & 1 \\ 
            &   -0.709647    & 2 \\ 
  $\dots$   &                &   \\ 
  2         &   -0.708318    & 2 \\ 
            &   -0.699671    & 2 \\ 
            &   -0.696110    & 1 \\ 
            &   -0.695246    & 1 \\ 
            &   -0.693130    & 2 \\ 
            &   -0.692557    & 1 \\ 
  $\dots$   &                &   \\ 
  3         &   -0.679517    & 1 \\ 
            &   -0.674428    & 1 \\ 
            &   -0.669761    & 2 \\ 
            &   -0.663414    & 2 \\ 
            &   -0.660698    & 1 \\ 
            &   -0.660617    & 2 \\ 
\end{tabular}              
\end{minipage}             
\hfill                     
\begin{minipage}[t]{4.5cm}

\center{{\large \bf s=$3\over2$}}
\vspace*{1em}

\begin{tabular}[t]{rcc}
\\
     $S$      &    $e_b$   & D \\
\\
\hline 
\\
  $1\over2$ &   -1.466517   & 2 \\ 
            &   -1.454860   & 1 \\ 
            &   -1.452973   & 2 \\ 
            &   -1.450302   & 2 \\ 
  $\dots$   &               &   \\ 
  $3\over2$ &   -1.463011   & 1 \\ 
            &   -1.455737   & 1 \\ 
            &   -1.451512   & 1 \\ 
            &   -1.448938   & 2 \\ 
            &   -1.446856   & 2 \\ 
            &   -1.446507   & 2 \\ 
  $\dots$   &               &   \\ 
  $5\over2$ &   -1.446507   & 2 \\ 
            &   -1.433347   & 1 \\ 
            &   -1.429590   & 2 \\ 
            &   -1.429420   & 1 \\ 
            &   -1.423490   & 1 \\ 
            &   -1.422074   & 1 \\ 
  $\dots$   &               &   \\ 
  $7\over2$ &   -1.419007   & 2 \\ 
            &   -1.402997   & 2 \\ 
            &   -1.396418   & 1 \\ 
            &   -1.394902   & 2 \\ 
            &   -1.394154   & 1 \\ 
            &   -1.391320   & 1 \\ 
\end{tabular}

\end{minipage}

\vspace*{1em}

\caption{Lowest part of the energy spectra for the N=15 Sierpinski gasket with 
$s={1\over2},1,{3\over2}$:  total spin $S$, energy per bond $e_b$ 
and degeneracy $D$ (the trivial Kramers degeneracy due to the z-component of
the total spin is not taken into account).
\label{ew}}
\end{minipage}
\ec
\end{table}

For \sez {\ } the complete spectrum can be calculated (see
\cite{voigt98jmmm}), but for {\sen} and {\sdz} the size of the matrix
which has to be diagonalized is much larger and only the lowest
eigen values in each subspace of total spin $S$ can be calculated. In
Table \ref{ew} we present the low-lying energies and their degeneracy
for the lowest four values of $S$. In general we present six eigen
values in each subspace of $S$, however with some exceptions:

\bi

\item For \sez {\ } and $S=1/2$ we give all eigen values below the first
eigen value with $S=3/2$.

\begin{figure}[ht]
\centerline{
\epsfig{file=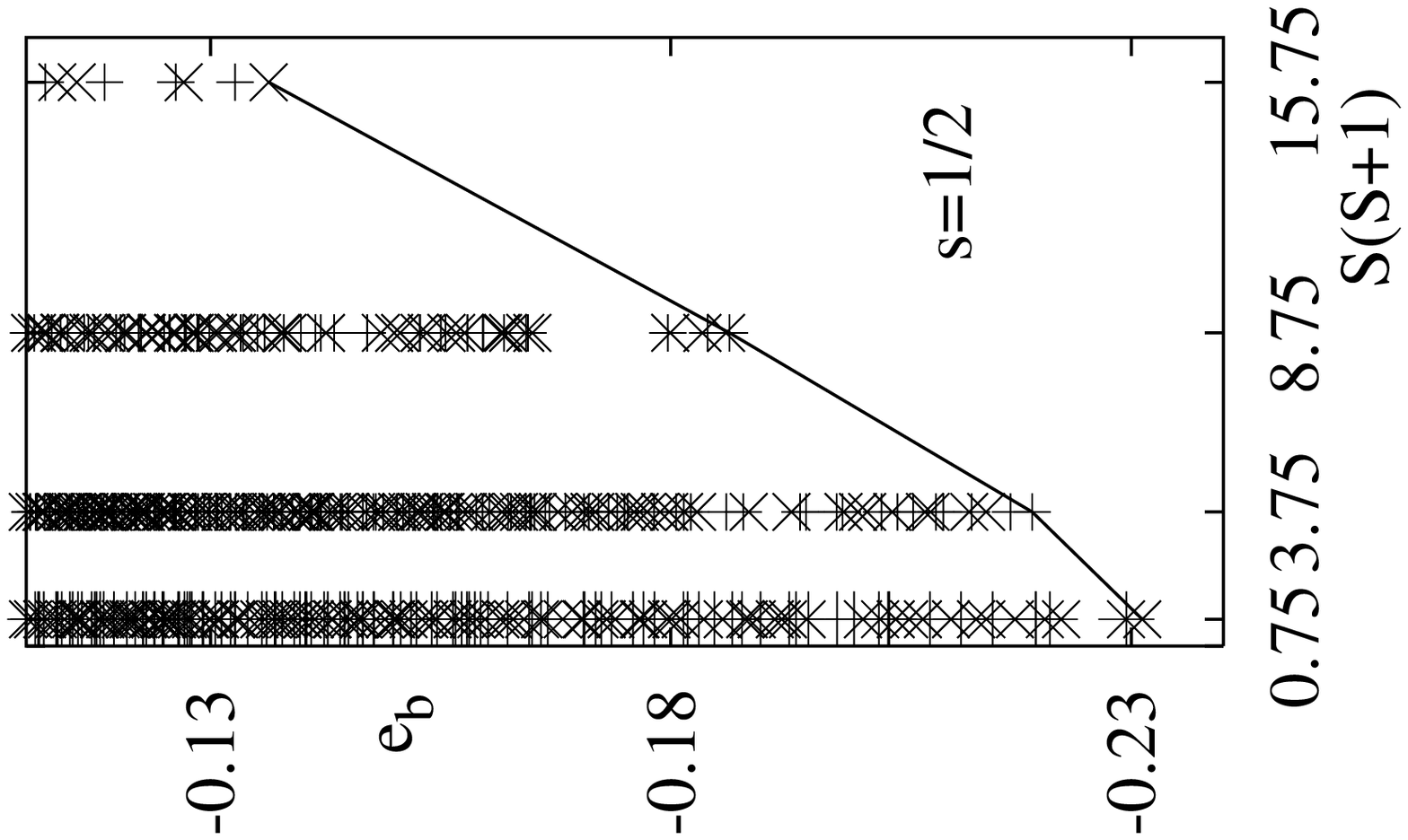,scale=0.45,angle=-90}
\hspace{-3em}
\epsfig{file=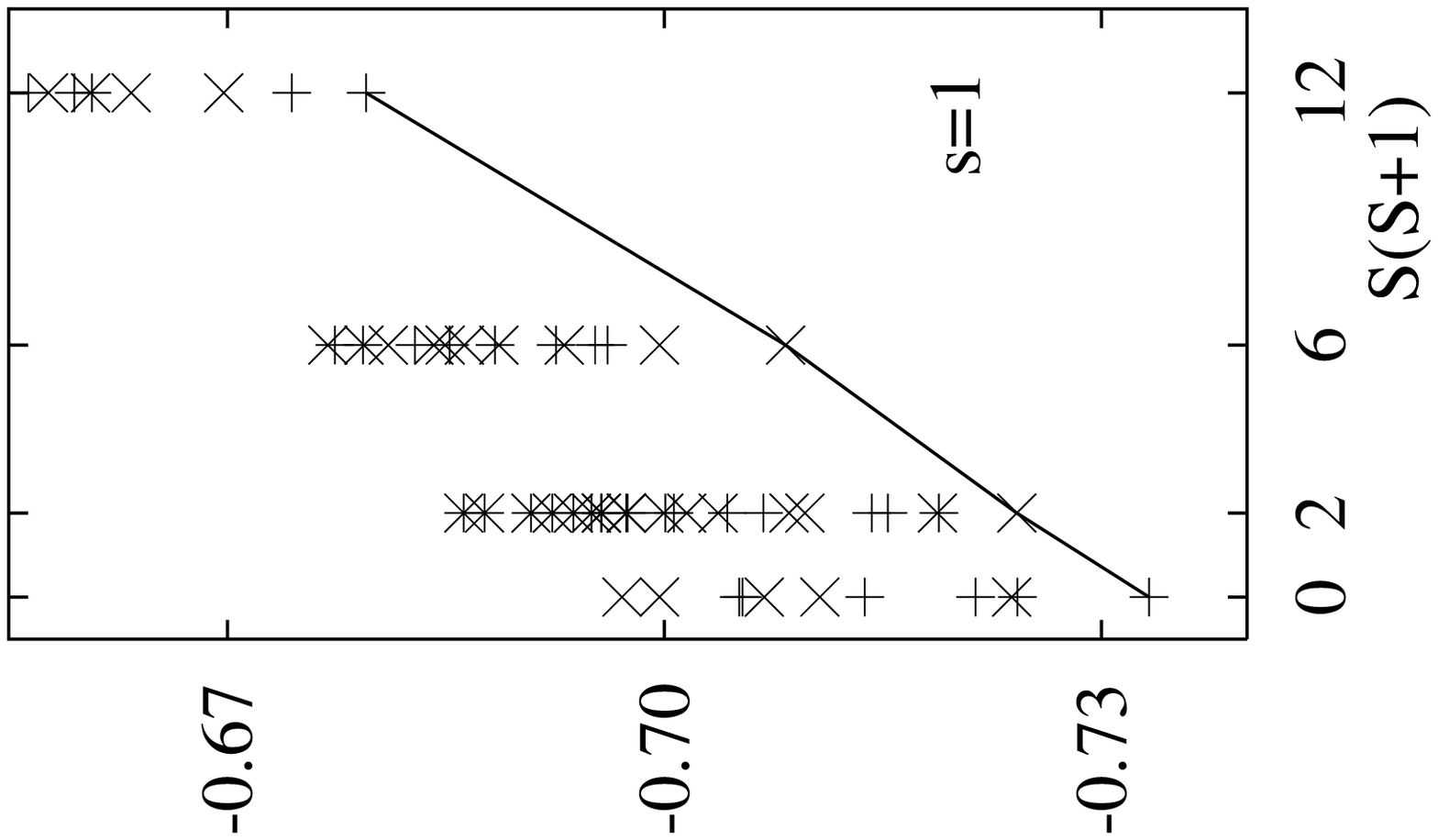, scale=0.45,angle=-90}
\hspace{-3em}
\epsfig{file=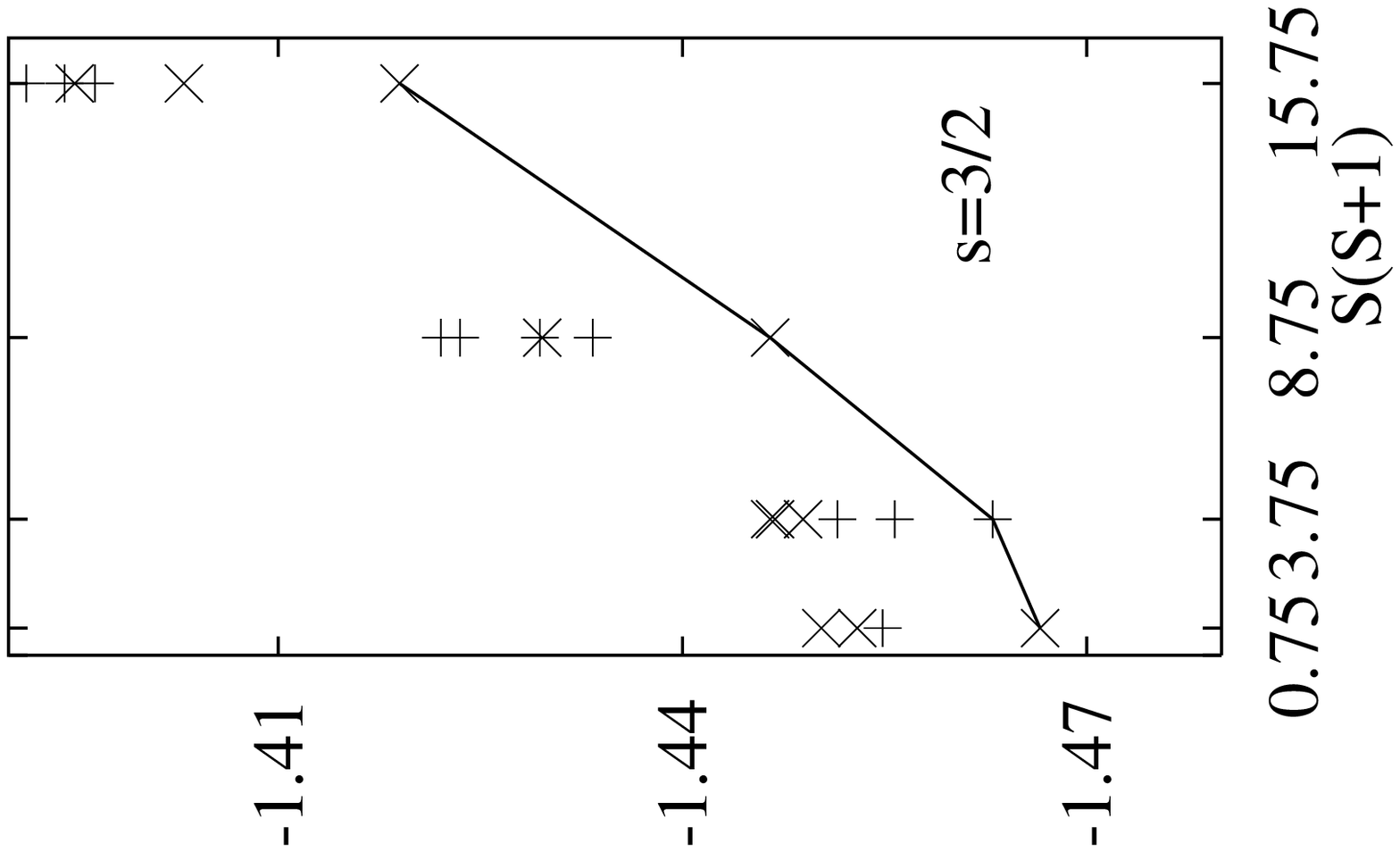,scale=0.45,angle=-90}}
\vspace*{1em}
\caption{The low-lying spectra of the Sierpi{\'n}ski 
gasket with N=15 vs. $S(S+1)$. The straight line connects the lowest
eigen values in every subspace of $S$. Non-degenerated eigen values are 
denoted by $+$, degenerated eigen values by $\times$.}
\label{fig_low_ew}
\end{figure}

\item For \sen {\ } we need already 1.7 million basis states to
calculate the singlet eigen states ($S=0$). Using the Lanczos procedure
with an extra projection onto the singlet subspace, we were able to
find the lowest 16 singlet eigen states.  In Table \ref{ew} we give
only the lowest 6 eigen values, but in the following calculation of
low-temperature thermodynamics all 16 eigen values will be used and all
of them are plotted in Fig.\ref{fig_low_ew}.

\item For \sdz {\ } we need already 97 million states in the $S=1/2$
subspace and large amount of computer power is necessary to calculate
the lowest 4 eigen states in this subspace.  In the subspaces with
higher $S$ we give the lowest 6 eigen values. 

\ei

Going back to a suggestion of Anderson \cite{anderson52}, which has
been recently picked-up in several papers
\cite{neuberger89,bernu92,bernu94,lhuillier00} the low-energy part of
the spectrum can be used to discuss the possibility of {\Ne} ordering
in the GS.  The lowest levels in each subspace up to $S \sim \sqrt{N}$
should be described by an effective Hamiltonian.
\be
H_{eff} = E_{0} + \frac{K}{N} {\vec S}^2 .
\label{level}
\ee
where $K$ is a constant for a given lattice.

These so-called quasi degenerate joint states collapse to a symmetry
broken {\Ne} state in the thermodynamic limit. Furthermore,
significantly above the family of quasi degenerate joint states a
second family of levels describing the magnon excitations are typical
for a HAF with {\Ne} ordering.

Indeed, a linear relation between the lowest eigen values $E_{min}(S)$
and $S(S+1)$ and a similar relation for the family of magnon
excitations has been observed in good approximation for the HAF on the 2d
square lattice \cite{neuberger89,richter94slo} as well as for the
triangular-lattice HAF \cite{bernu94}. The strong deviation from this
linear relation has been used as one argument that the HAF on the 2d
\kag lattice has a disordered ground state \cite{lecheminant97}. A
similar argumentation has been used already in \cite{voigt98jmmm} for the
{\sez} HAF on the Sierpi{\'n}ski gasket.

In Fig. \ref{fig_low_ew} we compare the spectra of the HAF on the
$N=15$ Sierpi{\'n}ski gasket with different spin quantum number $s$.

\vspace*{1em}

From Fig.\ref{fig_low_ew} it is obvious that for the Sierpi{\'n}ski
gasket there is a significant deviation from the linear relation between
$E_{min}(S)$ and $S(S+1)$, well seen for \sez {\ } and for \sdz {\ },
less pronounced for \sen. This deviation is much larger than in
corresponding {\Ne} ordered 2d systems with similar size (e.g. 
the square lattice).

Let us now discuss the excitations above the lowest states (see
Fig.\ref{fig_low_ew} and Table \ref{ew}). For \sez {\ } there is a
number of states with $S=1/2$ below the first state with $S=3/2$. In a
{\Ne} ordered system usually the first excitation is a state where the
quantum number $S$ of the total spin is enlarged by one (i.e.
$S=S_{GS}+1$), whereas the excitations with the same total spin as the
GS ($S=S_{GS}$) are quite far above the GS since they neither are
needed for the symmetry broken {\Ne} state nor for the magnons.  Hence
the existence of low-lying excitations with $S=S_{GS}$ are considered
as another indication for the lack of {\Ne} ordering
\cite{waldtmann98,lhuillier00}.  This property of the spectrum has been
observed for the \kag lattice, too \cite{waldtmann98}.

For the HAF on the Sierpi{\'n}ski gasket with higher spin quantum
number $s$ the situation seems to be different: For \sen {\ } only the
ground state and one singlet with $S=0$ exist below the first triplet
excitation with $S=1$. This second singlet is only slightly below the
first triplet excitation. In this context we note the recent paper of
Hida on the \kag lattice with \sen {\ } \cite{hida00}. The behavior
found there seems to be very similar to ours concerning the number of
the lowest singlets and the energy gap of these excitations. For \sdz
{\ } there is no other state between the ground state with $S=1/2$ and
the first excitation with $S=3/2$.

To be more specific concerning the deviation of the lowest excitations
from the effective model (\ref{level}) we have calculated the mean
square deviation $\delta$ defined by
\be
\delta  = \sqrt{\sum_{S=S_{GS}}^{S_0}\left (e_i(S) - 
[a \;+ \;b \;S(S+1)]\right)^2 }
\label{delta}
\ee
where $a$ and $b$ are determined to minimize $\delta$ and $e_i(S)$ are
the exact eigen values divided by the corresponding GS energy. Since
the linear-chain HAF has a critical GS (power-law decay of correlations
to zero) for half-integer $s$ and a disordered GS (exponential decay of
correlations to zero) for integer $s$ its corresponding $\delta_{LC}$
might be used as a criterion for the possibility of a disordered GS in
other lattices. Since $\delta$ is influenced by several factors as
system size, even or odd number of spins or possible {\Ne} ordering
with two or three sublattices we bear in mind that this criterion is
rough.

We present in Table \ref{sp_vgl_i} the quotient $\delta/\delta_{LC}$
for the lattices from Table \ref{eb_vgl}. 
\begin{table}
\renewcommand{\baselinestretch}{0.8}
\huge \normalsize \small
\bc
\begin{minipage}[t]{10cm}
\begin{tabular}[t]{lcccc}
 Lattice & N  & $s={1\over2}$ &  $s=1$ &  $s={3\over2}$\\
 \\
\hline\\ 
 Sierpinski gasket   & 15 & 1.870 & 0.821 & 12.981 \\
 honeycomb lattice   & 12 & 0.058 & 0.015 & 0.053  \\
 square lattice      & 10 & 0.078 & 0.017 & 0.026  \\
 \kag lattice        & 12 & 3.408 & 2.451 & 16.038 \\
 triangular lattice  & 12 & 1.543 & 0.147 & 0.239  \\ 
\end{tabular}
\end{minipage}
\ec
\vspace*{0.3cm}
\caption{Mean square deviation $\delta$ 
behavior (cf. eq.(\ref{delta})) scaled by the corresponding value
$\delta_{LC}$ of the
linear chain with $N=12$.
\label{sp_vgl_i}
}
\end{table}

For \sez {\ } the \kag lattice has the largest $\delta$ but the next in
the row is already the Sierpi{\'n}ski gasket. Both values are larger
than $\delta_{LC}$. The $\delta$ of the triangular lattice is smaller
than that of the \kag and the Sierpi{\'n}ski gasket but it is still
larger than $\delta_{LC}$, although the GS of the HAF on the triangular
lattice is {\Ne} ordered.
For the honeycomb and the square lattice (both have {\Ne} ordered GS)
$\delta$ is more than an order of magnitude smaller than in the other
lattices.
With increasing spin quantum number $s$ to $s=1$ the situation is
slightly changed.  The linear chain (the GS is now disordered not
critical) has a slightly higher value of $\delta$ than the
Sierpi{\'n}ski gasket. The \kag lattice is again the lattice with the
largest $\delta$, whereas all the other lattices have a $\delta$ which
is significantly smaller than that of the linear chain.
For \sdz {\ } one can see that both the \kag lattice and the
Sierpi{\'n}ski gasket have a $\delta$ which is one order of magnitude
larger than that of the linear chain, whereas the $\delta$ for the
other lattices is clearly below $\delta_{LC}$.
These results would favor a non-{\Ne} ordered GS for the Sierpi{\'n}ski
gasket for \sez {\ }, $s=1$ and \sdz {\ } and a close similarity to the
2d \kag lattice.

\section{Spin-spin correlations}

The spin-spin correlation $\langle {\vec s}_i  {\vec s}_{j} \rangle$ as
function of separation $r = |\vec{R}_i - \vec{R}_j | $ yields a direct
information on the magnetic ordering.  To discuss the possibility of
\Ne-like ordering in the GS the spin-spin correlation between the spins
within one sublattice (cf. e.g. all {\Large $\bullet$} in
Fig.\ref{fig1}) should be considered. Because the Sierpi{\'n}ski gasket
has no translational symmetry, the spin-spin correlation was averaged
over all pairs of spins with one and the same separation $r$. 
Furthermore the degeneracy of the GS for \sez {\ } and \sdz {\ } was
taken into account by an additional averaging over the two degenerated
states.

\begin{figure}[ht]
\centerline{
\epsfig{file=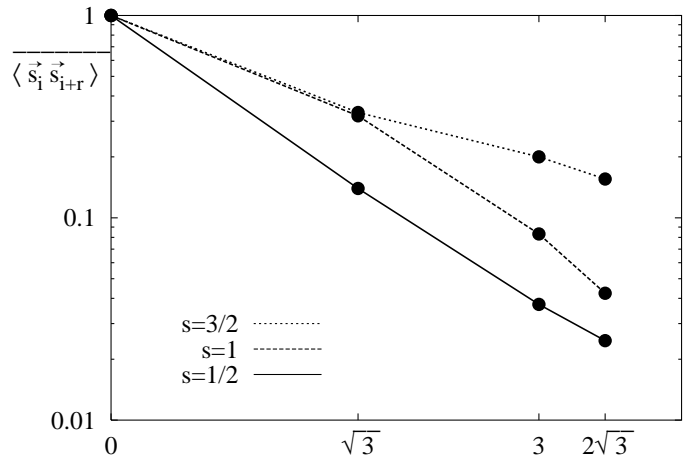,scale=0.4,angle=-90}}
%\vspace*{1em}
\caption{The scaled  spin-spin correlation $\overline{ \langle 
\vec{s}_i
\vec{s} _{j} \rangle }$ vs. distance 
$r = |\vec{R}_i - \vec{R}_j | $ for the Sierpi{\'n}ski 
gasket with N=15.
\label{fig_sisj_r}}
\end{figure}

In Fig. \ref{fig_sisj_r} the spin-spin correlation (scaled by the
on-site spin-spin correlation $\langle {\vec s}_i {\vec s}_{i} \rangle
= s(s+1)$) is shown in a semilogarithmic plot.
\vspace*{1em}

The spin-spin correlation shows a strong decay with increasing
distance. The decay is strongest for {\sez} but we do not see a
qualitative change in the behavior for $s=1$ and {\sdz}.  Though, the
presented correlations represent only a short length scale the linear
relation in the semilogarithmic plot suggests an exponential decay to
zero. We notice, that e.g. for the square lattice one has a significant
deviation from this behavior \cite{tomczak96co}.

\begin{figure}[ht] \centerline{
\epsfig{file=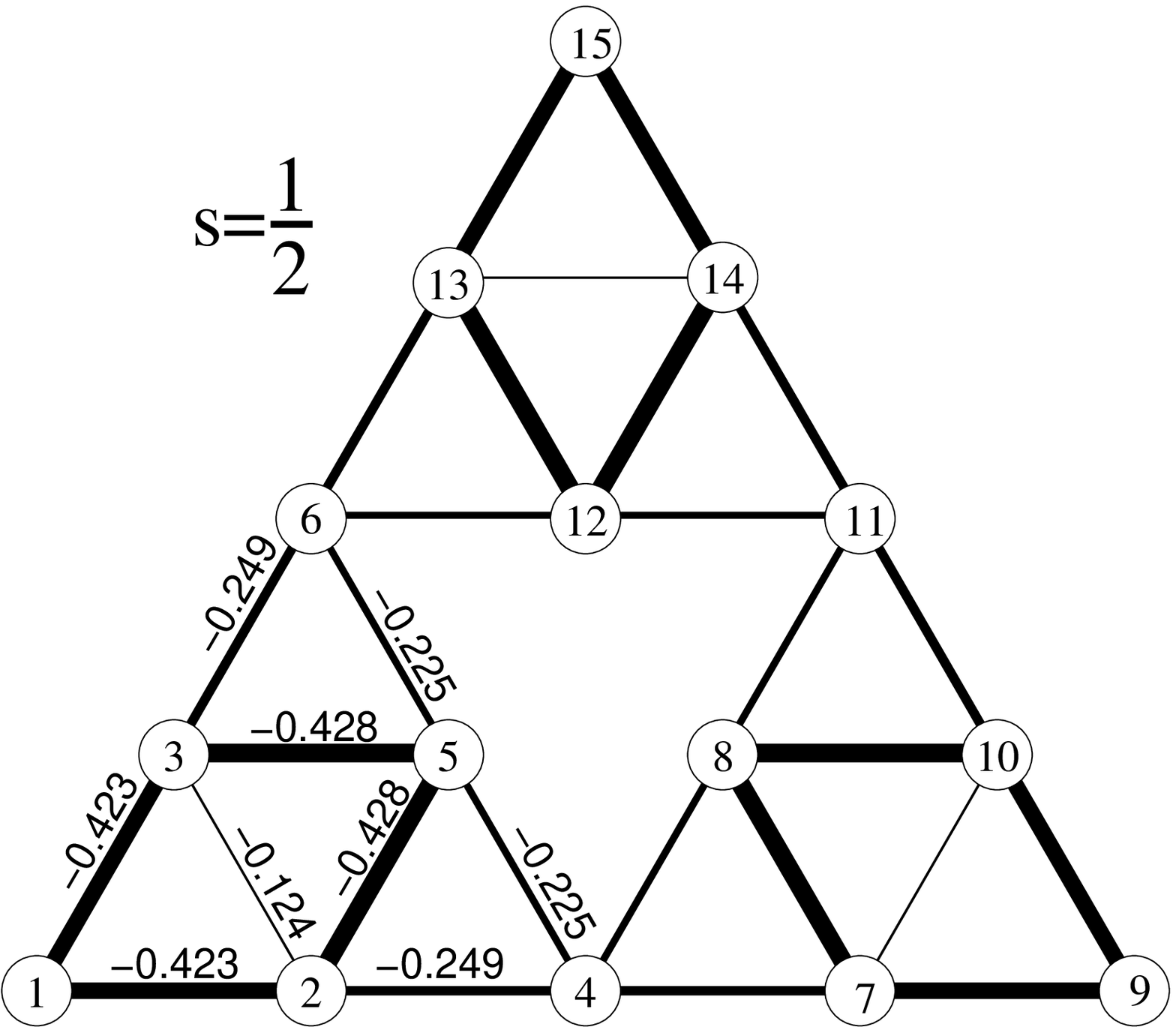,scale=0.45,angle=0} \hfill
\epsfig{file=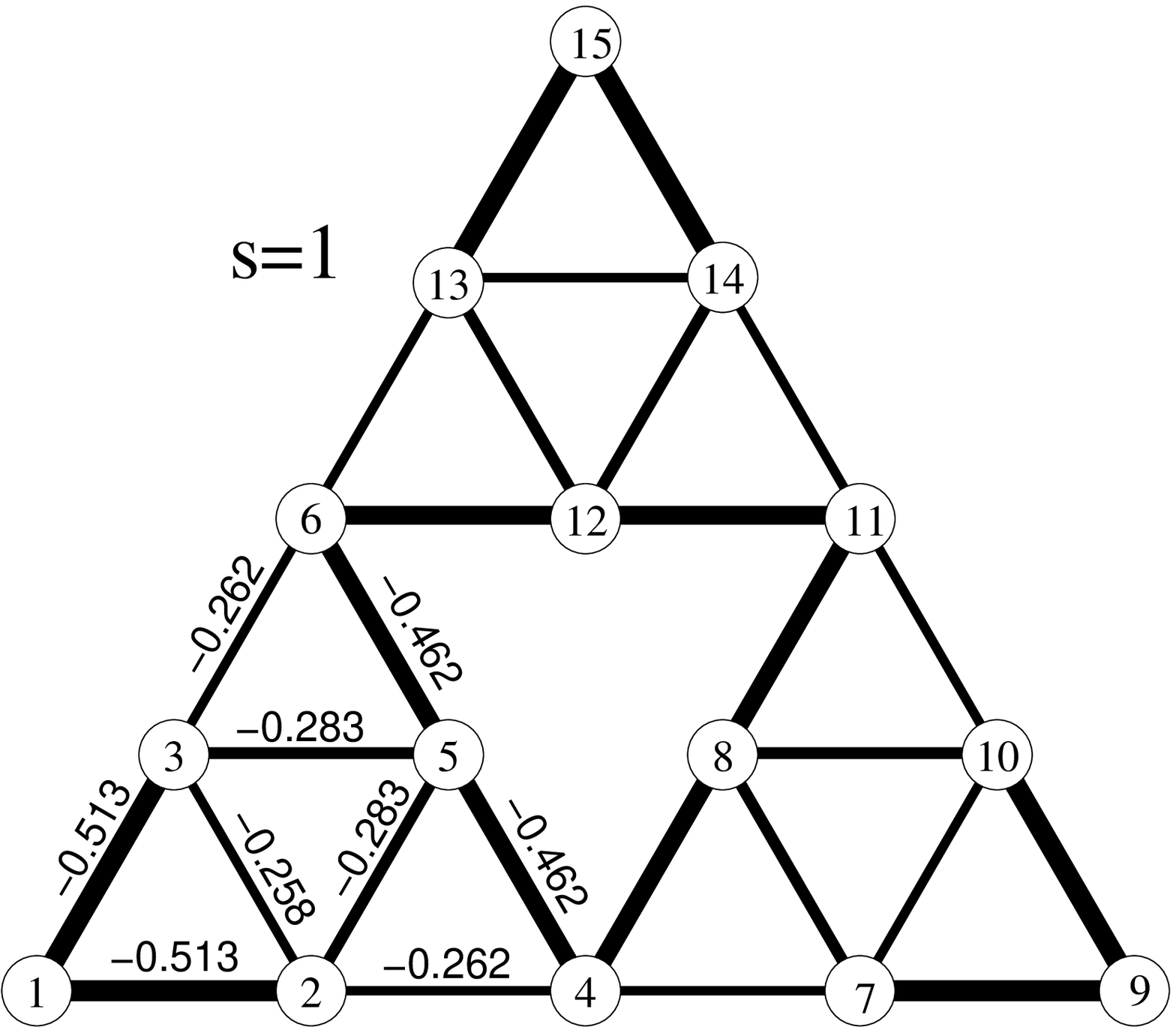,scale=0.45,angle=0} }

\centerline{
\epsfig{file=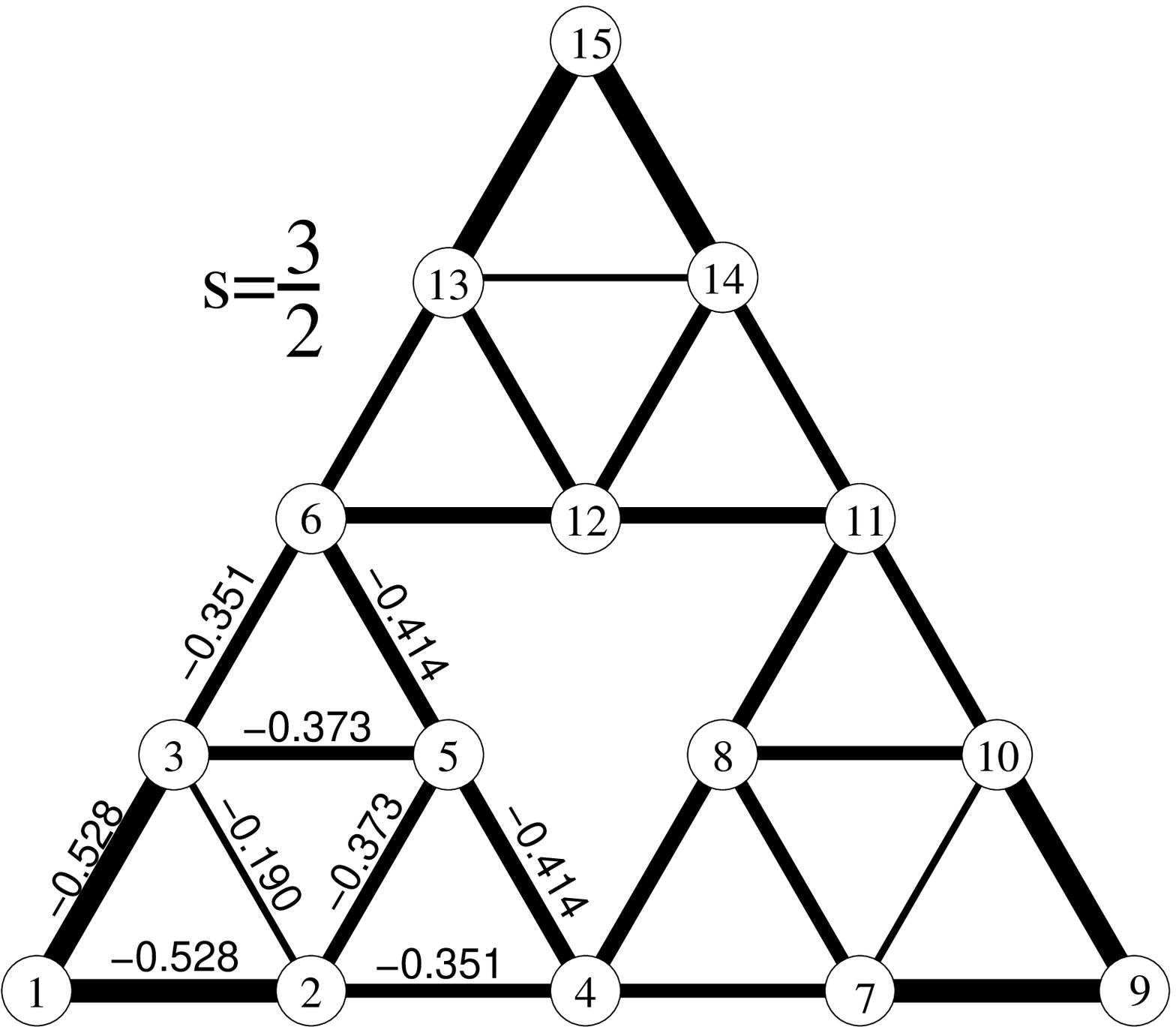,scale=0.45,angle=0}
}
\vspace*{.3cm}
\caption{The scaled nearest-neighbor 
spin-spin correlation in the ground state 
for the Sierpi{\'n}ski gasket with N=15.}
\label{fig_corr_sg15}
\end{figure}

As mentioned above the Sierpi{\'n}ski gasket is not translational
invariant and therefore, in principle, the spin-spin correlation
$\langle {\vec s}_i {\vec s}_{j} \rangle$ may depend not only on
distance $r = |\vec{R}_i - \vec{R}_j | $ but on both site indices $i$
and $j$.  To illustrate this in detail we show in Fig.
\ref{fig_corr_sg15} the value of the nearest-neighbor spin-spin
correlation scaled by the on-site spin-spin correlation $\langle {\vec
s}_i {\vec s}_{i} \rangle = s(s+1)$.  The width of the line corresponds
to the strength of the correlation.
Indeed, we see a strong variation in the  nearest-neighbor correlations
from bond to bond for all three values of $s$. This is a pure quantum
effect since for the classical GS (see Fig.\ref{fig1}) all
nearest-neighbor correlations are the same. We conclude, that the
interplay between quantum fluctuations and structural inhomogeneities
is important even in the model with the highest spin quantum number
{\sdz}.
For \sez {\ } it is interesting to note that a kind of plaquette
ordering occurs (cf. the correlations between lattice sites 1-2-5-3-1).
This kind of ordering does not appear in the lattices with higher spin
quantum number $s$.

\section{Low-temperature thermodynamics}

As argued in \cite{misguich99,sindzingre00,lhuillier00} the
low-temperature thermodynamics of a quantum spin system on a particular
lattice, 
especially the
specific heat and the entropy can be related to the magnetic ordering of
the system in the ground state.
For the \sez {\ } HAF on the Sierpi\'ski gasket such a relation between
low-temperature specific heat and magnetic order had already been
noticed \cite{voigt98jmmm}.  In \cite{sindzingre00} similar
considerations on the finite \kag lattice has been reported and the
authors conclude that the \kag lattice provides an example of a new
type of spin liquid with a non-magnetic excitation continuum adjacent
to the ground state. Following these considerations we will examine the
Sierpi{\'n}ski gasket with different spin quantum numbers $s$.
We will start with a comparison of exactly calculated low-temperature
thermodynamics of \sez {\ } systems. We compare a finite square lattice
(N=10) with the Sierpi{\'n}ski gasket with N=15.

\begin{figure}[ht]
\centerline{
\epsfig{file=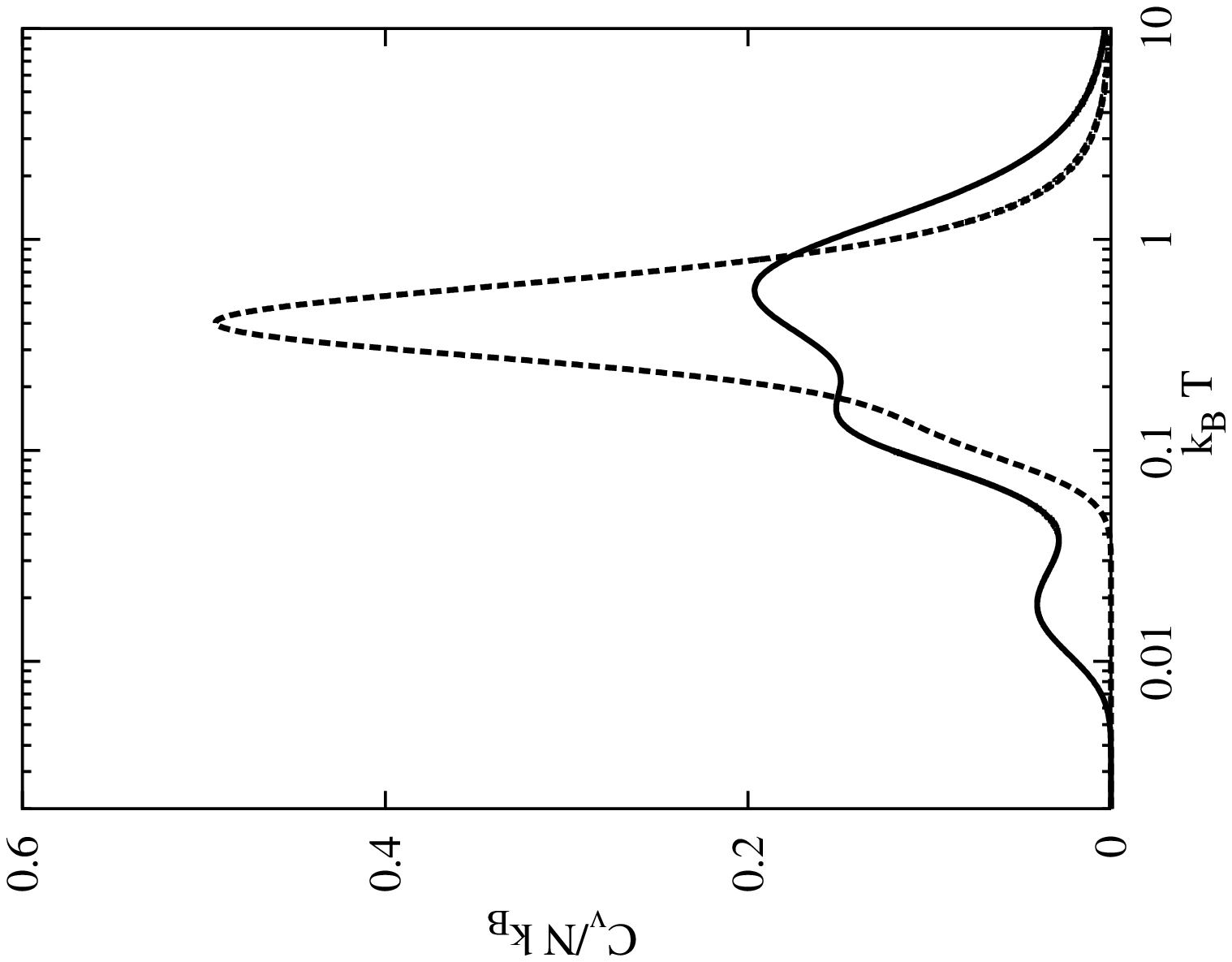,scale=0.5,angle=-90}
\hspace*{1em}
\epsfig{file=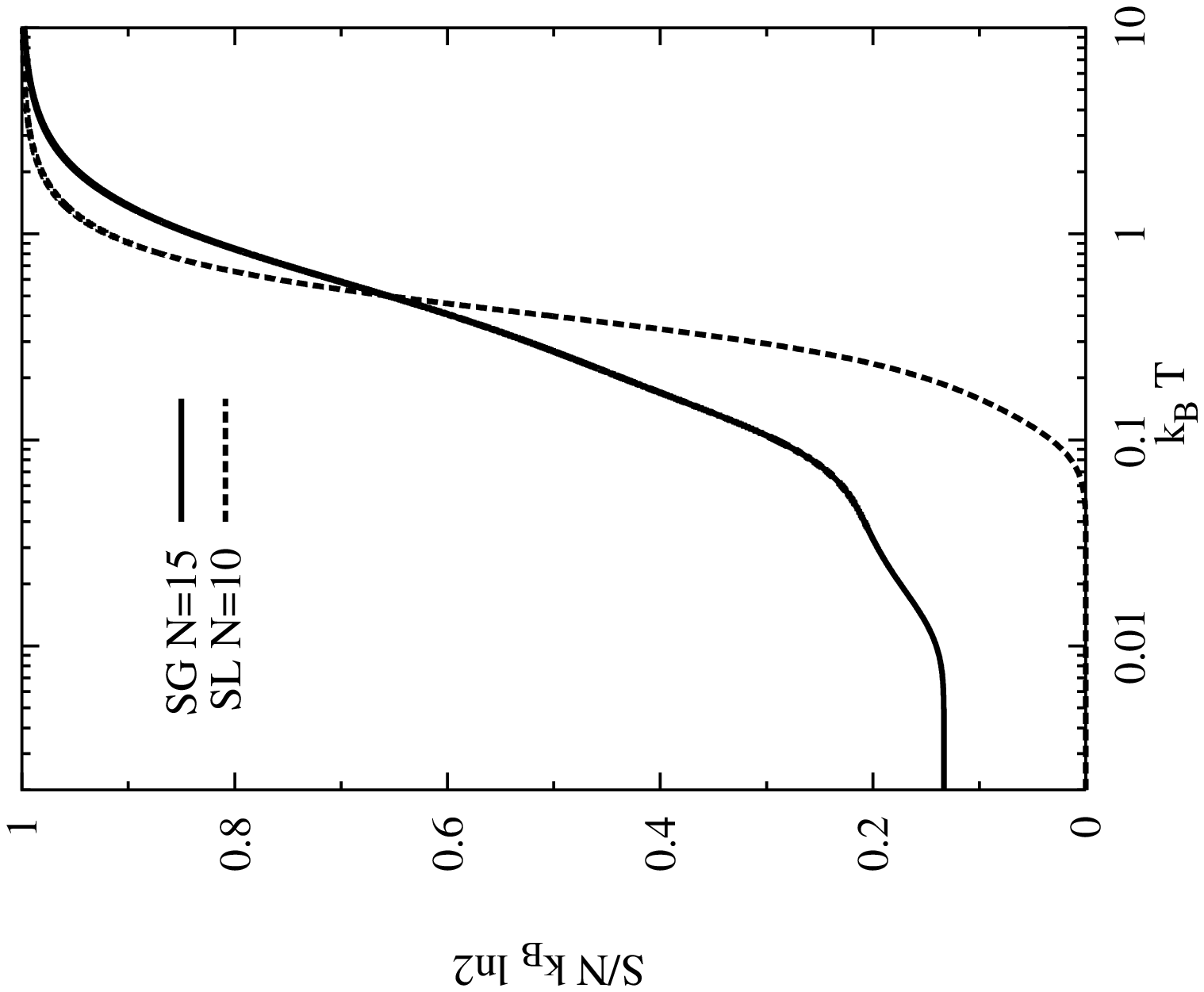,scale=0.5,angle=-90}}
\vspace*{1em}
\caption{Specific heat $C_v/N k_B$ (left panel) and  
entropy $S/N k_B ln2$ (right panel) of spin \sez {\ } systems: 
Sierpi{\'n}ski gasket  with N=15 (bold line) and  square lattice 
with N=10 (dashed line).
\label{s12_entropy}}
\end{figure}
\begin{figure}[ht]
\centerline{
\epsfig{file=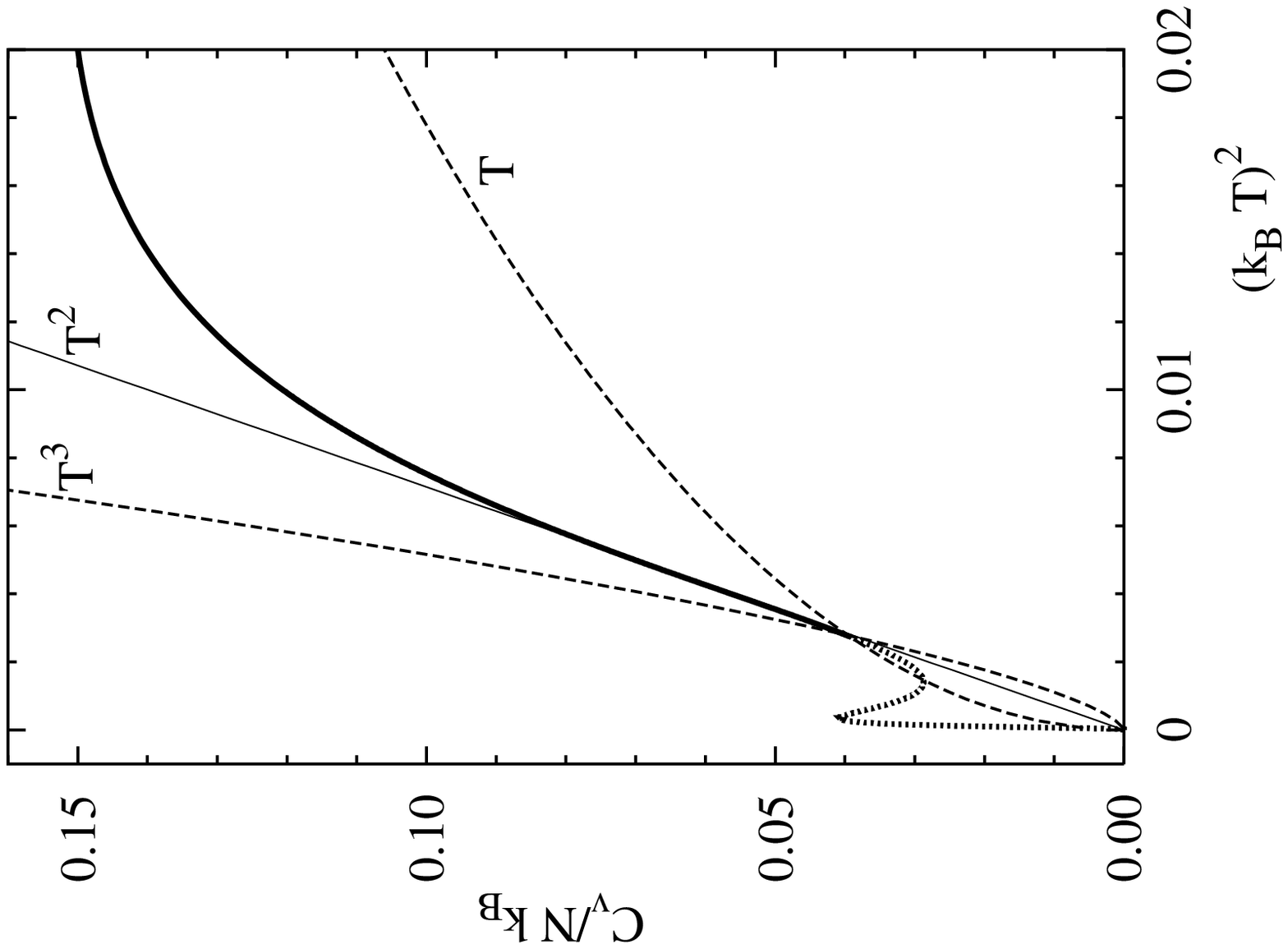,scale=0.5,angle=-90}
\hspace*{-1em}
\epsfig{file=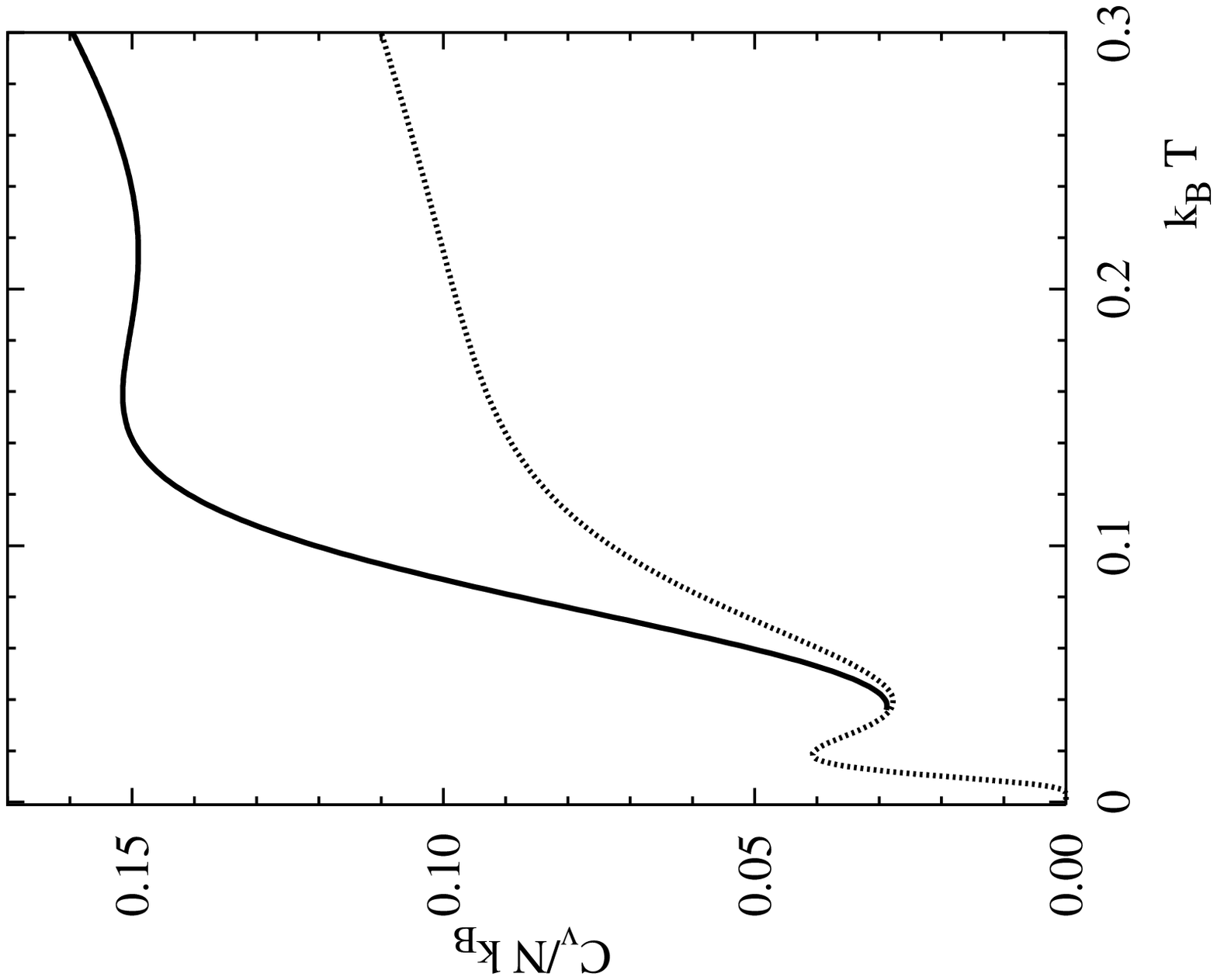,scale=0.5,angle=-90}}
\vspace*{1em}
\caption{Low-temperature specific heat $C_v$ of the \sez {\ } 
Sierpi{\'n}ski gasket with N=15 \newline 
left panel: $C_v$ is well fitted by a $T^2$ law  (light 
line) in the range from   $k_B T=0.05-0.09$. \newline 
right panel: Contribution of the doublet states with 
$S={1\over2}$ (dotted line) to the total specific heat.
\label{low_c_kt}}
\end{figure}

First we observe in both systems one peak in the specific heat slightly
above $k_B T = 0.5$. This peak is related to the energy scale beyond which
short-range correlations start to be broken. But in the Sierpinski
gasket there are two additional peaks below $k_B T = 0.2$ belonging to
low-energy scales: the first one at $k_B T = 0.018$ can be attributed
to the low-lying $S=1/2$ states (see right panel of Fig. \ref{low_c_kt}) 
and its position and even its existence may be strongly influenced by
finite-size effects, but the second peak at $k_B T = 0.16$ can be
related to the excitations with higher total spin $S>1/2$ (see right
panel of Fig.\ref{low_c_kt}) and in particular to the magnitude of the
spin gap $\Delta = E(S_{min}+1)-E(S_{min})$.
Comparing the entropy we also observe a different behavior. In the square
lattice there is about 85\% of the total entropy in the high-temperature
peak from $k_B T = 0.2 - \infty$. The total entropy in the Sierpi{\'n}ski
gasket is quite different from that. First we find a non-zero entropy at
$k_B T = 0$, which corresponds to the degeneracy of the GS. Second we
observe that already 45\% of the total entropy belong to the first and
second peak in the specific heat for $k_B T = 0 - 0.2$. Only the remaining
55\% of the total entropy is contained in the high-temperature peak of the
specific heat. This behavior of the Sierpi{\'n}ski gasket is quite similar
to the observations for the \kag lattice \cite{sindzingre00}.

It should also be noted that $C_v$ does not decrease exponentially
below the first excitation with $S=3/2$ (i.e. within the spin gap).
This behavior can be seen in the left panel of Fig. \ref{low_c_kt}.
Below the peak at $k_B T=0.16$ the specific heat is well fitted by a
$T^2$ law in the range of $k_B T =0.05-0.09$. This non-exponential
behavior of $C_v$ is related to the presence of low-lying $S=1/2$-
states within the spin gap. However, as can be seen in the right panel
of Fig. \ref{low_c_kt}, to some extent higher spin channels do
contribute also to $C_v$ in this range.

The observed $T^2$ behavior of the specific heat and the additional
low-tempera/-ture maximum are very similar to the \sez {\ } \kag
lattice \cite{sindzingre00}. It suggests a close relation between the
low-temperature physics of the HAF on the Sierpi{\'n}ski gasket and on
the \kag lattice.  Assuming the same scenario as for the \kag lattice
the spin gap (and hence the second peak in $C_v$) probably survives in
the thermodynamic limit, but this spin gap is filled by a continuum of
states having the same total spin $S$ as the GS (and hence the first
peak in $C_v$ at $k_B T = 0.018$ disappears for $N \to \infty$).  As
recently pointed out by C.Lhuillier and coworkers
\cite{sindzingre00,lhuillier00} for this kind of low-energy spectrum,
we have (in contrast to the power-law $T^2$ decay of $C_v$) a thermally
activated susceptibility which decays exponentially within the spin
gap.

\begin{figure}[ht]
\centerline{
\epsfig{file=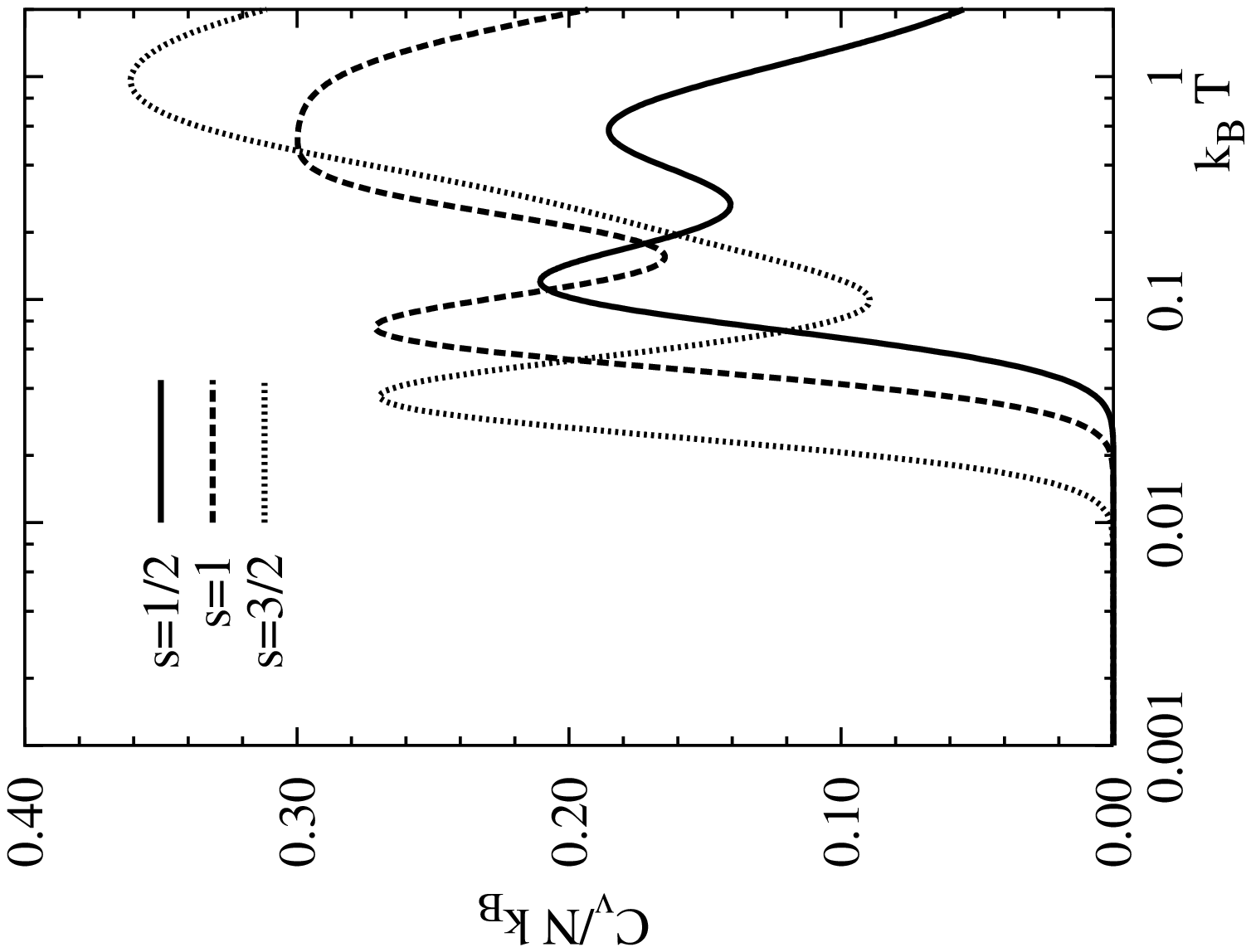,scale=0.5,angle=-90}
\hspace*{1em} 
\epsfig{file=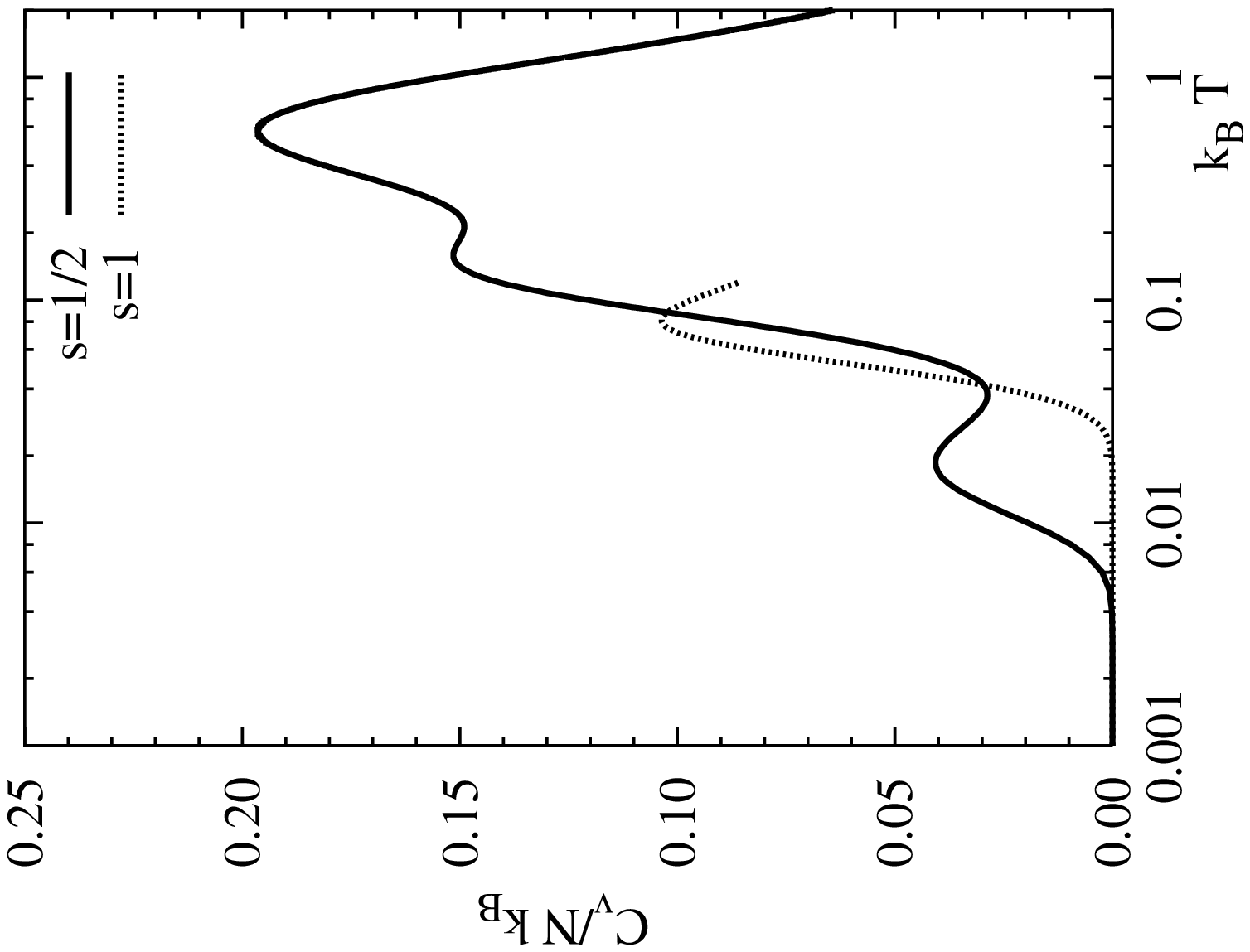,scale=0.5,
angle=-90}}
\vspace*{1em}
\caption{Comparison of low-temperature specific heat for Sierpi{\'n}ski
gasket with N=6 (left panel) and N=15 (right panel)}
\label{c_vgl}
\end{figure}

For the Sierpi{\'n}ski gasket with N=6 the complete thermodynamics is
numerically accessible and for a comparison we show in the left panel
of Fig. \ref{c_vgl} the specific heat data for this system for \sez,
\sen {\ } and \sdz. For the Sierpi{\'n}ski gasket with N=15 only the
data for \sez {\ } can be calculated exactly. However, for \sen {\ } we
are able to obtain the specific heat for low temperatures $k_B T < 0.1$
by using all computed low-lying excitations. The two data sets for N=15
are shown in the right panel of Fig. \ref{c_vgl}.

From the figure it can be seen that for the system with N=6 one
observes one additional low-temperature peak in the specific heat. This
property does not depend on the spin quantum number $s$, only the height
and the position of this first peak changes. These findings for $N=6$
support our argumentation for the N=15 system with {\sez}. The lowest
peak in $C_v$ in this system at $k_B T = 0.018$ may be attributed to a
finite-size effect, whereas the other additional peak at $k_B T = 0.16$
is a real physical effect.

The calculation of the same data for N=15 is exact only for \sez. For
\sen \ we can calculate only a number of lowest energies (see section
\ref{ener}) and therefore reliable results for the specific heat can be
obtained for temperatures $k_B T<0.1$ only.  But this temperature range
is sufficient to demonstrate that (i) there is no peak at very low
temperature and no $T^2$ behavior (notice, that in difference to
{\sez} there is no set of singlet states within the spin gap) and that
(ii) the second low-temperature peak exists also in this case but it is
shifted to lower temperatures (notice that the magnitude of the spin
gap is smaller for $s=1$ than for {\sez}).  The high-temperature peak
for \sen \ cannot be displayed, since the corresponding energy scale is
beyond the temperatures we are able to consider.
 
For \sdz {\ } we  cannot give comparable results, the calculated
low excitations can provide specific-heat data only for $k_B T<0.05$. In
this temperature range we did not observe any peculiarities in the
specific heat therefore the data are not shown.

\section{Conclusion}

The results of a numerical investigation of the ground state and low
temperature thermodynamics of the Heisenberg antiferromagnet on the
finite Sierpi\'nski gasket with N=15 are presented. These data enable
us to draw conclusions with respect to the magnetic ordering of the
spin system. This is done by comparing them with the corresponding data
of the one-dimensional chain and the two-dimensional square,
triangular, honeycomb and \kag lattices. For these systems the magnetic
ordering is well understood.  We argue that the quantum Heisenberg
antiferromagnet on the Sierpi{\'n}ski gasket may remain disordered not
only for \sez, but also for \sen {\ } and \sdz. The data suggest that
in contrast to the one-dimensional Heisenberg antiferromagnet there is
no fundamental distinction between half-integer and integer spin
quantum number $s$.  In conclusion we find a close correspondence
between the physics of the quantum Heisenberg antiferromagnet on the
Sierpi{\'n}ski gasket and on the \kag lattice. Assuming a short range
magnetic order in both systems we relate this correspondence to the
similar local geometry of both lattices with the same coordination
number $z=4$ and the frustration.

\section{Acknowledgment}

A.V.\ is grateful to the Institute of Molecular Physics of the Polish
Academy of Sciences in Pozna{\'n} (Poland) for kind hospitality during
his stay. This work was supported by the DFG under Grant-Nr. Ri 615/5-1
and KBN Grant-Nr. PO3B 046 14.

\end{document}